\documentclass{iau} 
\usepackage{graphicx}

\title[Spectral flattening of Crab Giant Pulses] 
{Spectral Flattening of Crab Giant Pulses at Low Frequencies}

\author[B. W. Meyers et al.]   
{B.~W.~Meyers$^{1,2,3}$,
S.~E.~Tremblay$^{1,2}$,
N.~D.~R.~Bhat$^{1,2}$
\and
R.~M.~Shannon$^{1,3}$
}

\affiliation{
$^1$International Centre for Radio Astronomy Research (ICRAR), Curtin University\\
	1 Turner Avenue, Technology Park,	Bentley, 6102, WA, Australia\\ 
	email: {\tt bradley.meyers@postgrad.curtin.edu.au} 
\\[\affilskip]
$^2$ARC Centre of Excellence for All-Sky Astrophysics (CAASTRO)
\\[\affilskip]
$^3$CSIRO Astronomy and Space Science, Australia Telescope National Facility\\ 
	P.O. Box 76, Epping, NSW 1710, Australia
}

\pubyear{2017}
\volume{337}  
\jname{Pulsar Astrophysics -­ The Next 50 Years}
\editors{P. Weltevrede, B.B.P. Perera, L. Levin Preston \& S. Sanidas, eds.}

\begin{document}

\maketitle

\begin{abstract}
The frequency dependence of normal pulsar radio emission is typically observed to be a power law, with some indications of a flattening or turnover at low frequencies ($\lesssim100$\,MHz). 
The spectrum of the Crab pulsar's giant pulse emission has not been examined as closely. 
We conducted simultaneous wideband observations of the Crab pulsar, with the Parkes radio telescope and the Murchison Widefield Array, to study the spectral behaviour of its giant pulses. 
Our analysis shows that the mean spectral index of Crab giant pulses flattens at low frequencies, from $ -2.6\pm0.5 $ between the Parkes bands, to $ -0.7\pm1.4 $ between the lowest MWA subbands.
\end{abstract}

\firstsection 
\section{Introduction}
The Crab pulsar emits giant pulses (GPs) with energetics orders of magnitude greater than the underlying ``normal'' emission (e.g. \cite{cordes2004}).
These GPs appear to be broadband in the radio regime (e.g. \cite{hankins2007}; \cite{hankins2015}), and are detectable across the entire electromagnetic spectrum.
In general, the spectral behaviour of Crab GPs over a wide frequency range has not been studied in depth (though see \cite{mikami2016}). 
For normal pulsars, one expects a power law behaviour with a spectral index of $\langle\alpha\rangle = -1.41\pm0.96$ (\cite{bates2013}), while some pulsars exhibit a flattening or turn-over at low frequencies (e.g. \cite{maron2000}).
One can then ask: Do Crab GP spectra follow a single power law, even at low frequencies?

To answer this question, we obtained simultaneous observations with the Murchison Widefield Array (MWA) and the Parkes radio telescope, spanning 117--3600\,MHz.
\cite[Meyers \etal\ (2017)]{meyers2017} show that at low frequencies, the Crab GP spectrum is poorly described by a single power law as defined at higher frequencies (i.e. the spectrum is flattening).

\section{Overview}
Within the single pulse sensitivity limits of both instruments, we are able to cross-match individual GPs across our observed frequency range (sampled in three non-contiguous subbands from the MWA and two from Parkes).
The number of pulses detected in each subband ranged from 90 (at 120\,MHz) to $>6000$ (at 732\,MHz) in one hour of observations. 
Of these, six main pulses and one interpulse were matched across all five subbands.

\begin{figure}
\centering
\includegraphics[width=2.9in]{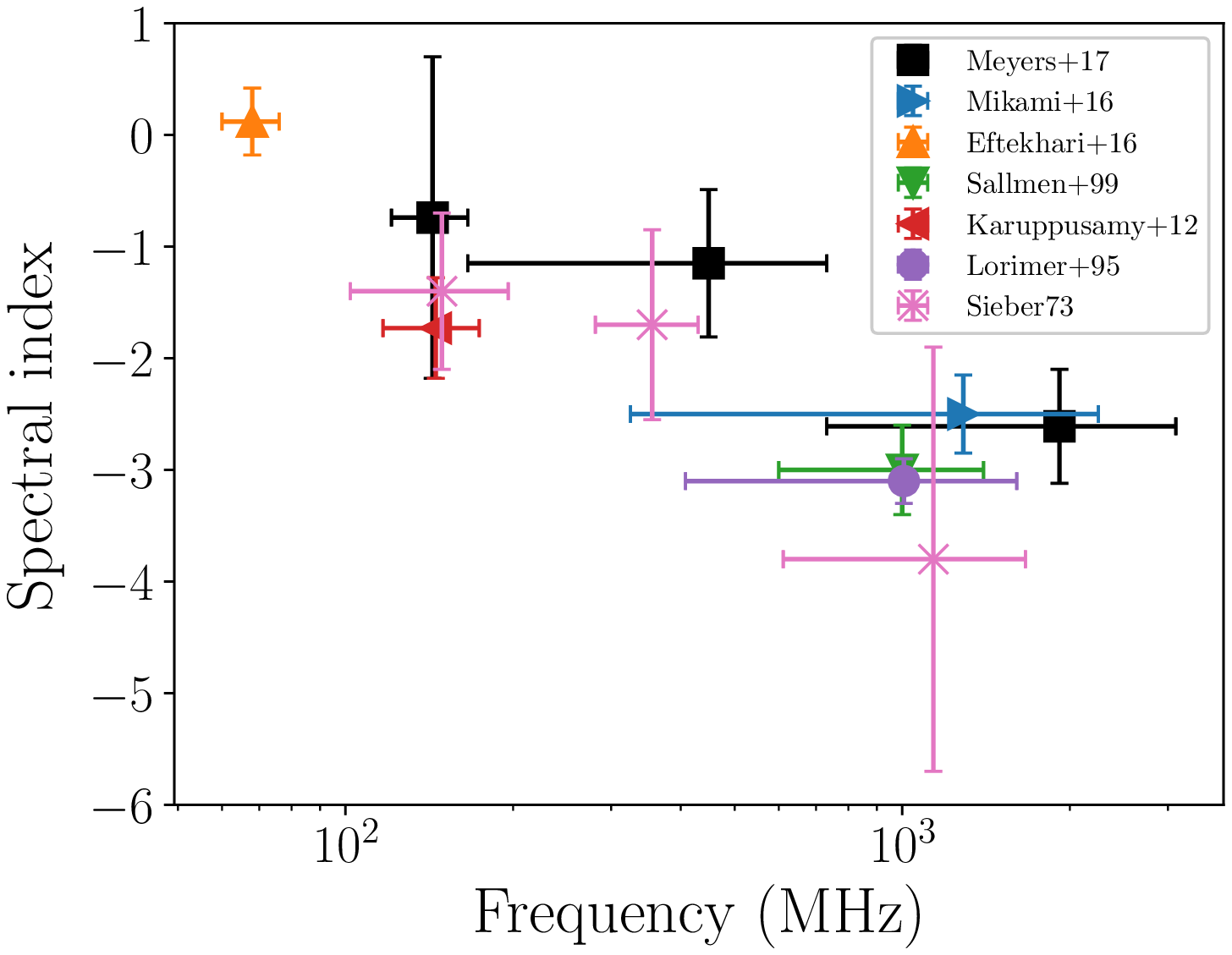}
\caption{Spectral indices vs. frequency, highlighting the flattening spectrum of Crab giant pulses at low frequencies. References:\,\cite{meyers2017},\,\cite{mikami2016}, \cite{eftekhari2016},\,\cite{sallmen1999},\,\cite{karuppusamy2012},\,\cite{lorimer1995},\,\cite{sieber1973}.}
\label{fig1}
\end{figure}

For every GP with two or more simultaneous subband detections, we calculated the spectral index (see \cite{meyers2017} for details). 
The spectral index distributions evolved from a relatively narrow, steep distribution between the two Parkes subbands ($ \alpha=-2.6\pm0.5 $ between 732 and 3100\,MHz), to a much wider, shallower distribution between the lowest MWA subbands ($ \alpha=-0.7\pm1.4 $ between 120.96 and 165.76\,MHz).
In the context of other spectral index measurements in the literature (see Figure~\ref{fig1}), the observed spectral flattening over a wide frequency range is reinforced.

\section{Implications}
Current models for GP radio emission do not make any explicit predictions for such a spectral flattening. 
Whether this spectral behaviour is an intrinsic aspect of the emission, or a consequence of some magnetospheric propagation effects, remains unclear.
This result also implies that perhaps certain models of fast radio bursts (FRBs), particularly those that involve GP emission, are less favourable (see \cite{meyers2017}).
Clearly, this result highlights the importance of simultaneous wideband studies of other GP emitting pulsars (e.g. PSR J1937+2134) in order to determine whether the Crab is a special case, or if this spectral behaviour is typical of GP emission.

\end{document}